\documentclass[fleqn,usenatbib]{mnras}

\usepackage{newtxtext,newtxmath}

\usepackage[T1]{fontenc}  
\usepackage{ae,aecompl}

\usepackage{graphicx}	
\usepackage{float}
\usepackage{amsmath}	
\usepackage{braket}
\usepackage{lastpage}
\usepackage{cleveref}
\usepackage{placeins} 

\usepackage{caption}
\usepackage{multirow}

\title[]{A Unified Maxwell–Bloch Framework for Multi-periodic 6.7 GHz Methanol Flaring in G9.62+0.20E}

\author[Rashidi et al.]{T. Rashidi,$^{1}$ V. Anari,$^{1,2}$ O. Powles,$^{1}$ G. C. MacLeod,$^{3,4}$ Y. Tanabe,$^{5,6}$ 
\newauthor Y. Yonekura,$^{5}$ F. Rajabi$^{1}$\thanks{E-mail: rajabf1@mcmaster.ca}   
\\
$^{1}$Department of Physics and Astronomy, McMaster University, 1280 Main Street West
Hamilton, Ontario, L8S 4M1, Canada \\
$^{2}$Department of Physics and Astronomy, The University of Western Ontario, 1151 Richmond Street, London, Ontario N6A 3K7, Canada \\
$^{3}$The Open University of Tanzania, P.O. Box 23409, Dar-Es-Salaam, Tanzania \\
$^{4}$Hartebeesthoek Radio Astronomy Observatory, PO Box 443, Krugersdorp, 1741, South Africa \\
X6001, Potchefstroom 2520, South Africa \\
$^{5}$Center for Astronomy, Ibaraki University, 2-1-1 Bunkyo, Mito, Ibaraki 310-8512, Japan \\
$^{6}$The Research Institute for Time Studies, Yamaguchi University, 1677-1 Yoshida, Yamaguchi-city, Yamaguchi 753-8511, Japan.\\
}

\date{}

\begin{document}
\label{firstpage}
\pagerange{\pageref{firstpage}--\pageref{lastpage}}
\maketitle

\begin{abstract}
We analyze a decade of 6.7 GHz methanol monitoring data in G9.62+0.20E, 
confirming the known periodicities of $p_{1} = 241.3 \pm 2.3$~d and 
$p_{2} = 52.5 \pm 0.3$~d, and identifying three new cycles at 
$p_{3} = 127.0 \pm 1.6$~d, $p_{4} = 163.9 \pm 2.9$~d, and $p_{5} = 204.1 \pm 1.5$~d. 
The 241.3-d and 204.1-d periods occur in multiple velocity channels, while the others are confined to single components. Despite their diverse morphologies and timescales, all flares can be reproduced within a unified Maxwell–Bloch framework operating in the fast-transient superradiance regime, driven by narrow periodic pump excitations. 
Model fits yield consistent environmental parameters across periodicities 
(temperatures, collisional timescales), pointing to broadly uniform 
physical conditions in the masing region. The discovery of new periodicities 
and their unified Maxwell-Bloch modelling provide a consistent picture of 
multi-periodic flaring in G9.62+0.20E and support superradiance as a general 
framework for maser flaring.

\end{abstract}




\section{Introduction}\label{sec:Introduction}

The brightest known Galactic methanol maser source, G9.62+0.20E, is a massive star-forming region associated with an ultracompact \( \mathrm{H_{II}} \) region and young massive stars. It is part of a star-forming complex located in the Near 3 kpc Arm \citep{Hofner1996}, at a parallax distance of \( 5.2_{-0.6}^{+0.6}\) kpc from the Sun \citep{Sanna2009}. The source lies at the edge of a molecular cloud core and shows strong free–free radio emission at centimeter wavelengths \citep{Garay1993, Hofner1996, Testi2000}.

Beyond its continuum properties, G9.62+0.20E is remarkable for hosting multiple masing species that make it a benchmark source for variability studies. These include class~II methanol masers at 6.7~GHz \citep{Menten1991} and 12.2~GHz \citep{Batrla1987,Caswell1995}, hydroxyl masers at 1665 and 1667~MHz \citep{Caswell1998}, and water maser at 22~GHz \citep{Hofner&Churchwell1996}. Long-term monitoring with HartRAO revealed periodic flaring of several methanol and hydroxyl maser features with a dominant period of about 243~d \citep{Goedhart2003, Goedhart2005, VanderWalt2009, Goedhard2019}. In addition, a shorter secondary period of \(52.1 \pm 0.3\)~d was identified in the 6.7~GHz methanol transition, in a single velocity component at \(v_{\mathrm{lsr}} = 8.8~\mathrm{km~s}^{-1}\) \citep{MacLeod2022}.

Mid-infrared continuum emission has also been detected toward G9.62+0.20E \citep{DeBuizer2003}, and near-infrared observations revealed two sources located close to the water and hydroxyl maser regions \citep{Testi1998a}. These detections are significant because class II methanol masers are thought to be radiatively pumped by infrared photons.

The variability of the 6.7~GHz methanol masers in G9.62+0.20E was first investigated by \citet{Goedhart2004}, following their earlier report of periodicity in this source \citep{Goedhart2003}. Although the physical origin of the variability remains uncertain, classical explanations within maser theory generally fall into two categories: (i) disturbances propagating through the masing region and (ii) variations in the flux of seed or pump photons. For G9.62+0.20E, the disturbance picture is disfavoured because the flux densities return to nearly the same level between successive flares \citep{VanderWalt2009}. Within the seed photon-flux framework, several specific mechanisms have been proposed, including colliding-wind binaries \citep{VanderWalt2009}, periodic accretion in a young binary system \citep{Araya2010}, pulsational instabilities of young accreting stars \citep{Inayoshi2013}, and spiral shocks induced by a young binary orbiting within a circumbinary disc \citep{Parfenov2014}.  

Extending beyond models that attribute variability to periodic pump modulation within the maser picture, \citet{Rajabi2023} showed that the observed flares can arise from the medium’s cooperative response to periodic excitation within the transient superradiance regime. In that study, \citet{Rajabi2023} employed the Maxwell–Bloch equations (MBEs) to establish a unified framework connecting the maser and superradiant regimes, demonstrating that the 243-d flares in G9.62+0.20E can arise naturally from cooperative radiative dynamics within the transient regime. Following the subsequent discovery of the secondary $\sim$52-d periodicity, \citet{MacLeod2022} investigated whether it could be interpreted as a secondary maximum (``ringing”) of the 243-d superradiant solution proposed by \citet{Rajabi2023}. The predicted ringing maxima, however, did not align with the observed 52-d flares, effectively ruling out the 52-d periodicity as the ringing of the 243-d superradiant solution. Instead, they suggested that the shorter periodicity may originate from either a secondary pulsation mode or a distinct periodically varying source within the same region, with continued monitoring required to discriminate between these possibilities.  

Although the ringing scenario was excluded, this does not preclude superradiance itself. In the present work, we extend the MBE modelling framework to all identified periodicities in G9.62+0.20E and show that the flaring behaviour across different periods and morphologies can be consistently reproduced within the transient superradiance regime. Rather than interpreting the secondary and newly identified periods as artefacts of the 243-d cycle, we model each flare as an independent periodicity driven by narrow pump pulses whose period closely matches the observed cycle.  

Superradiance is a cooperative spontaneous emission process that arises when coherence is established in an ensemble of population inverted atoms or molecules. In extended systems, emitters couple through a common electromagnetic field and radiate cooperatively, much more efficiently than they would independently \citep{Dicke1954}. Within the framework of the MBEs, superradiance and maser action emerge as two complementary regimes of radiation: masers correspond to a slow, quasi-steady state, while superradiance highlights a fast-transient regime. In \citet{Rajabi2016a, Rajabi2016b, Rajabi2017, Rajabi2019, Rajabi2020}, it was shown that superradiance can reproduce both the sharp rise followed by a slow decay seen in many flares, as well as more symmetric flare profiles, making it a natural framework for modelling variability observed in maser-hosting regions.

The structure of this paper is as follows. In Sec.~\ref{sec:dataANDframework}, we provide an overview of the observational data, followed by a description of our model. The results are presented in Sec.~\ref{sec:results} and discussed in Sec.~\ref{sec:discussion}. The Maxwell–Bloch system of equations is reviewed in Appendix~\ref{subsec:MBEs}, and the period analysis is described in Appendix~\ref{subsec:periodogram-analysis}. Finally, Appendix~\ref{subsec:Parameters} presents the detailed results of MBE modelling for flares across five periods and 11 velocity channels.

\section{DATA and TheoreticalFramework}\label{sec:dataANDframework} 
\subsection{Data}\label{subsec:data}

The 6.7~GHz methanol maser data analysed in this work were obtained with the Hitachi 32-m radio telescope, accessible through the Ibaraki methanol maser monitoring database (iMet\footnote{\url{https://vlbi.sci.ibaraki.ac.jp/iMet/}}). Monitoring of G9.62+0.20E began on 30 December 2012, with observations typically carried out once per day. The sensitivity of the program corresponds to a $1\sigma$ noise level of 0.3~Jy. At 6.7~GHz, the Hitachi telescope has an aperture efficiency of 55–75\%, a beam size of $4^\circ 6$, and side-lobe levels of 3–4\%.  

In this study, we used all data collected in this monitoring campaign up to 29 June 2022. Further details of the telescope and observing setup are given by \citet{Yonekura2016}.

\subsection{Theoretical modelling framework}\label{subsec:modelframework}  

To model the multi-periodic flaring light curves of G9.62+0.20E, we employ MBEs (see Appendix~\ref{subsec:MBEs}). These equations are a set of coupled differential equations that describe the interaction between light and matter by following the evolution of the population inversion density, the induced polarization, and the radiation field \citep{Gross1982, Benedict1996}. This framework has been successfully used to reproduce both periodic \citep{Rajabi2023, Rashidi2025} and non-periodic \citep{Rajabi2019, Houde2019} flares in maser-hosting regions. In particular, \citet{Rajabi2023} modelled periodic methanol and hydroxyl flares in G9.62+0.20E, while \citet{Rashidi2025} applied the same framework to periodic methanol flares in G22.356+0.066.

In a system of emitters (i.e. inverted atoms or molecules), coherence develops on a characteristic timescale $T_R$, defined as
\begin{equation} \label{eq:T_R}
    T_R = \frac{8 \pi}{3 \lambda^2 (nL)} \, \tau_{\mathrm{sp}},
\end{equation}
where $nL$ is the inverted column density, with $n$ denoting the population inversion density and $L$ the length of the system, $\lambda$ the transition wavelength, and $\tau_{\mathrm{sp}}$ the spontaneous emission timescale of a single atom or molecule for that transition \citep{Rajabi2016b, Rajabi2017}.  

The onset of superradiance depends not only on $T_R$ but also on competing decoherence processes. Collisions and other random interactions can disrupt coherence, and these effects are described in the MBEs by two characteristic timescales: $T_1$, which represents population inversion relaxation through non-coherent processes such as inelastic collisions, and $T_2$, which accounts for dephasing due to random processes such as elastic collisions.  

For single-burst superradiance, cooperative emission occurs when $T_R < T_1, T_2$, ensuring that coherence is established before it is destroyed. In the case of periodic flaring, however, the definition of $T_R$ becomes more complex, as it depends on the interplay between the pumping timescale $T_{\mathrm{P}}$ and the evolution timescale of coherence. Following \citet{Rajabi2023}, we therefore define an initial $T_{R,0}$, determined prior to the onset of the first modelled flare from the corresponding inverted column density. This $T_{R,0}$ provides a relevant timescale for interpreting the MBE solutions and for assessing the condition $T_{R,0} < T_1, T_2$.

The superradiance timescale, $T_R$, together with $T_2$, sets a critical threshold for the inverted population column density. When this threshold is exceeded, the system transitions from the quasi-steady maser regime to the fast-transient superradiance regime. Increases in the pumping rate or in the background continuum can trigger this transition by raising the inverted column density $nL$. In the fast-transient regime, coherence builds rapidly and the system produces intense bursts of radiation \citep{Benedict1996, Rajabi2020, Rajabi2023}.

As shown by \citet{Feld1980} (see also \citealt{Rajabi2020}), the MBEs exhibit two limiting regimes. In the quasi-steady maser regime, the population inversion and polarization evolve on timescales longer than the relaxation and dephasing timescales ($T_1$ and $T_2$), so the radiation field adjusts to the pump and the response follows its profile closely. In contrast, when the evolution occurs on timescales shorter than $T_1$ and $T_2$, the system enters the transient superradiance regime, producing bursts rather than tracking the pump. In this regime, the flare morphology is largely decoupled from the detailed pump profile: short, symmetric pump pulses can give rise to long-duration flares with either symmetric or asymmetric shapes. Consequently, reproducing the observed flares does not require the pump to mirror the data. In our modelling, we therefore match only the pump \emph{period} to the observations and drive the system with narrow symmetric pulses. Using broadly consistent environmental parameters ($T_1$, $T_2$), this framework provides a unified basis for modelling flare profiles across velocity channels and periodicities in G9.62+0.20E.

\section{Results}\label{sec:results}
\subsection{Multi-periodicity of G9.62+0.20E}

We examined the light curves of all available velocity components to search for periodic variability and to confirm the two previously reported periods. Period analysis was carried out using Lomb–Scargle (LS) periodograms (see Appendix~\ref{subsec:periodogram-analysis}), and the candidate periods were further verified by phase-folding \citep{Stellingwerf1978} and the Jurkevich method \citep{Jurkevich1971}. This analysis confirmed the primary period of \(p_{1} = 241.3 \pm 2.3\)~d and recovered the secondary period \(p_{2} = 52.5 \pm 0.3\)~d, consistent with the values previously reported by \citet{MacLeod2022} in the \(v_{\mathrm{lsr}} = 8.8\)~km~s\(^{-1}\) channel. In addition, we identified three new periodicities: (1) \(p_{3} = 127.0 \pm 1.6\)~d, observed only at \(v_{\mathrm{lsr}} = -0.2\)~km~s\(^{-1}\), (2) \(p_{4} = 163.9 \pm 2.9\)~d, also restricted to the \(v_{\mathrm{lsr}} = -0.2\)~km~s\(^{-1}\) channel, and (3) \(p_{5} = 204.1 \pm 1.5\)~d, which is present across eight velocity channels spanning \(v_{\mathrm{lsr}} = -0.8\) to \(6.5\)~km~s\(^{-1}\) (see Table~\ref{tab:periods_summary}).  

It is worth noting that \citet{Goedhart2014} reported a period of 131.0~d in the \(v_{\mathrm{lsr}} = -0.2\)~km~s\(^{-1}\) channel, which is close to the \(p_{3}\) period reported here at the same velocity. As summarized in Table~\ref{tab:periods_summary}, the features exhibiting \(p_{2}\), \(p_{3}\), and \(p_{4}\) also flare at the primary period \(p_{1}\), whereas this does not necessarily hold for those flaring at \(p_{5}\).

\begin{figure}
    \centering
    \includegraphics[width=1.0\linewidth]{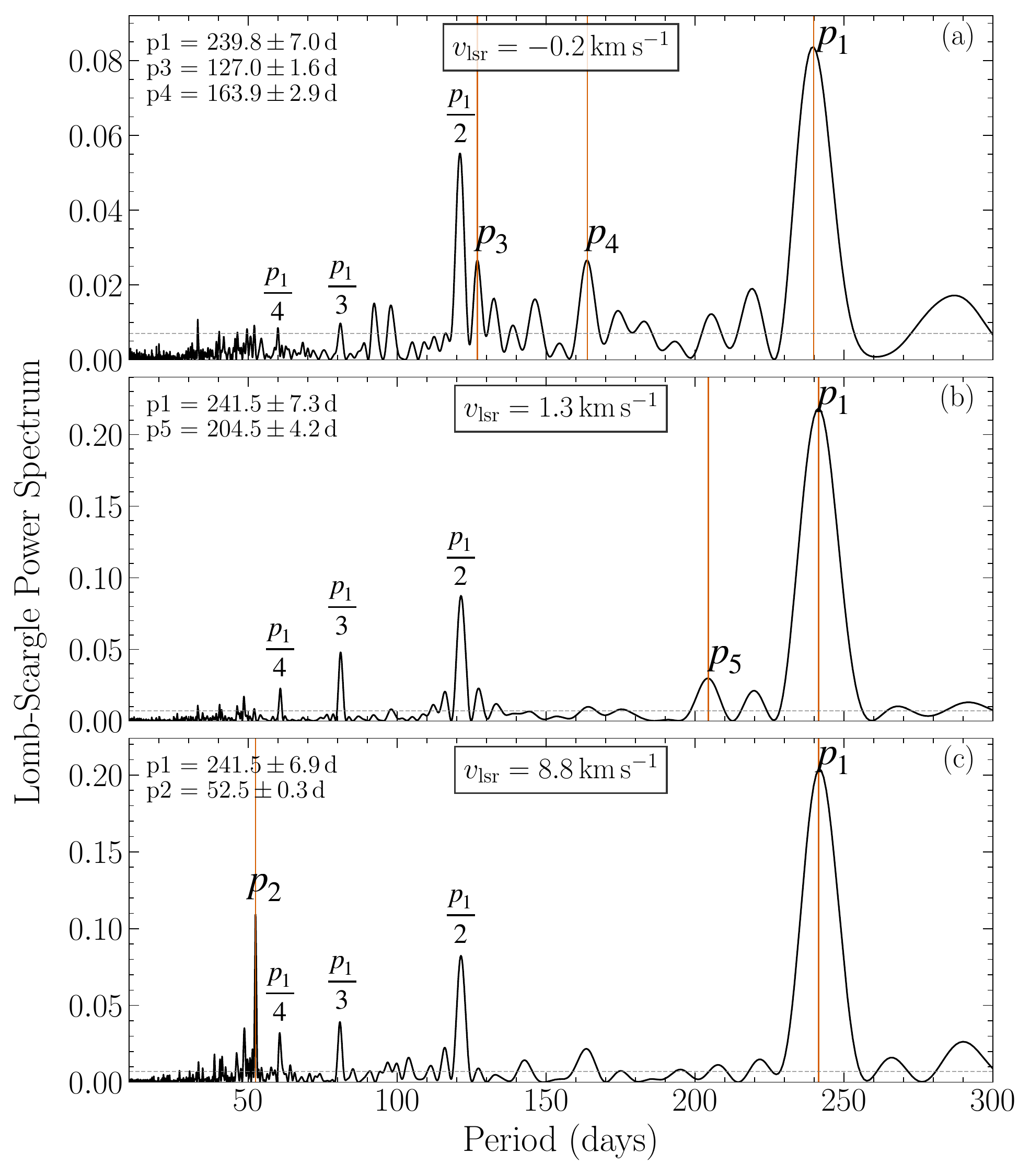}
    \caption{Lomb--Scargle periodograms for three selected velocity components: \textbf{\textit{(a)}} \(v_{\mathrm{lsr}} = -0.2\)~km~s\(^{-1}\), \textbf{\textit{(b)}} \(v_{\mathrm{lsr}} = 1.3\)~km~s\(^{-1}\), and \textbf{\textit{(c)}} \(v_{\mathrm{lsr}} = 8.8\)~km~s\(^{-1}\), along with the exact period values derived from the periodograms. For each velocity channel, the most prominent peaks are labelled, together with the first three harmonics of the strongest peak. The dashed grey line shows the significance threshold, defined as the average of the maximum powers from 1000 Lomb--Scargle periodograms generated from simulated light curves with flux values randomly distributed between $-0.3$ and $+0.3$~Jy, representing the observational noise level.}   
    \label{fig:All periodograms}
\end{figure}

To illustrate the detection process, LS periodograms for three representative velocity components are shown in Fig.~\ref{fig:All periodograms}. The most prominent peaks are labelled, together with the harmonics of the primary period. The dashed horizontal line at a power level of 0.007 indicates the significance threshold. This threshold was determined by generating 1000 simulated light curves with flux values randomly drawn from the observational noise range of $\pm 0.3$~Jy and calculating their LS periodograms. The average of the maximum powers from these simulations defines the threshold, such that peaks above this line are considered significant, while those below it are consistent with noise. For peaks above the significance threshold, additional analysis was performed to determine whether they correspond to genuine signals or artifacts introduced by the finite time window of the LS analysis. These cases were further examined using the CLEAN algorithm to mitigate sampling-related artifacts (see Appendix~\ref{subsec:periodogram-analysis}).  

We note that none of the newly identified secondary periods are harmonically related to \(p_{1}\). Although additional peaks above the significance threshold appear in some periodograms (Fig.~\ref{fig:All periodograms}), we focused on the most prominent ones for MBE modelling.

\begin{table*}
\centering
\makebox[\textwidth]{\parbox{0.9\textwidth}{
\caption{Periods detected in the 6.7~GHz methanol maser light curves of G9.62+0.20E. 
The first column lists the velocity channel (\(v_{\mathrm{lsr}}\)), and the subsequent columns give the detected periodicities: \(p_{1}\) (primary cycle), \(p_{2}\) (secondary cycle), \(p_{3}\), \(p_{4}\), and \(p_{5}\). The periods are derived from Lomb--Scargle (LS) periodograms and verified using phase folding and the Jurkevich method. All periods are listed in days.}
\label{tab:periods_summary}
}}
\begin{tabular}{ 
|p{2cm}||p{2.5cm}||p{2.5cm}||p{2.5cm}||p{2.5cm}||p{2.5cm}|}
\hline
\( v_{\mathrm{lsr}}~(\mathrm{km~s}^{-1}) \) & \( p_1 \)~(d) & \( p_2 \)~(d) & \( p_3 \)~(d) & \( p_4 \)~(d) & \( p_5 \)~(d) \\
\hline
\hline
$-0.8$ &  \(-\) & \(-\) & \(-\) & \(-\) & \( 202.7 \pm 3.9 \)  \\

$-0.2$  & \( 239.8 \pm 7.0 \) & \(-\) & \( 127.0 \pm 1.6 \) & \( 163.9 \pm 2.9 \) & \(-\) \\

0.4 & \( 242.4 \pm 6.8 \) & \(-\) & \(-\) & \(-\) & \( 205.1 \pm 4.2 \) \\

1.3 & \( 241.5 \pm 7.3 \) & \(-\) & \(-\) & \(-\) & \( 204.5 \pm 4.2 \) \\
3.3 & \( 241.5 \pm 7.4 \) & \(-\) & \(-\) & \(-\) & \( 204.5 \pm 10.7 \)
\\
4.1 & \( 238.2 \pm 7.2 \) & \(-\) & \(-\) & \(-\) & \( 203.9 \pm 4.6 \) \\
5.0 & \( 243.2 \pm 7.0 \) & \(-\) & \(-\) & \(-\) & \( 205.7 \pm 4.4 \)
\\
5.4 & \(-\) & \(-\) & \(-\) &  \(-\) & \( 202.1 \pm 4.2 \) \\
6.5 & \(-\) & \(-\) & \(-\) &  \(-\) & \( 203.9 \pm 4.6 \) \\
8.1 & \( 243.2 \pm 7.0 \) & \(-\) & \(-\) & \(-\) &  \(-\) \\
8.8 & \( 241.5 \pm 6.9 \) & \(  52.5 \pm 0.3 \) & \(-\) & \(-\) & \(-\) \\
\hline
\end{tabular}
\end{table*}

\subsection{Maxwell-Bloch modelling} \label{subsection: MBE modeling}

For each identified period, the light curves were phase-folded to reduce noise and construct an average flare profile. The phase axis was then converted to time, yielding a representative flare shape as a function of time.

The MBEs were applied to these average profiles using a periodic pump to sustain the population inversion in the system:
\begin{equation} \label{eq:pump}
    \Lambda_n(z, \tau) = \Lambda_0 + \Lambda_{1}\sum_{m=0}^{\infty} 
    \frac{1}{\cosh^2{\left[ \left( \tau - \tau_0 - m\tau_1\right) / T_{\mathrm{P}} \right]}}.
\end{equation}
Here, $\Lambda_0$ and $\Lambda_1$ denote the amplitudes of the constant quiescent pump and the periodic pump pulses, respectively. In contrast to \citet{Rajabi2023} and \citet{Rashidi2025}, who allowed the pump-pulse amplitudes to vary within a given cycle, we adopted a constant amplitude to model all flares within each cycle. This approach reflects our focus on modelling average flare profiles for each period rather than individual events and was applied consistently to all periodicities.

The pump period, $\tau_1$, was set equal to the average value extracted
from the periodograms. As
stated in Sec.~\ref{subsec:modelframework}, narrow pump pulses of
a few days were used for different periods. A delay time, $\tau_0$,
was introduced to align the model with the data.

In all simulations, the initial population inversion density corresponds to a level of $\sim 0.1$~cm$^{-3}$ across a velocity range of 1~km~s$^{-1}$. The non-coherent relaxation and dephasing timescales were set to average values of $T_{1} \simeq 38.7$~d and $T_{2} \simeq 5.3$~d, with variations of approximately $\pm 9\%$ across different velocity channels and periodicities.

\begin{figure}
    \centering
\includegraphics[width=1.0\linewidth]{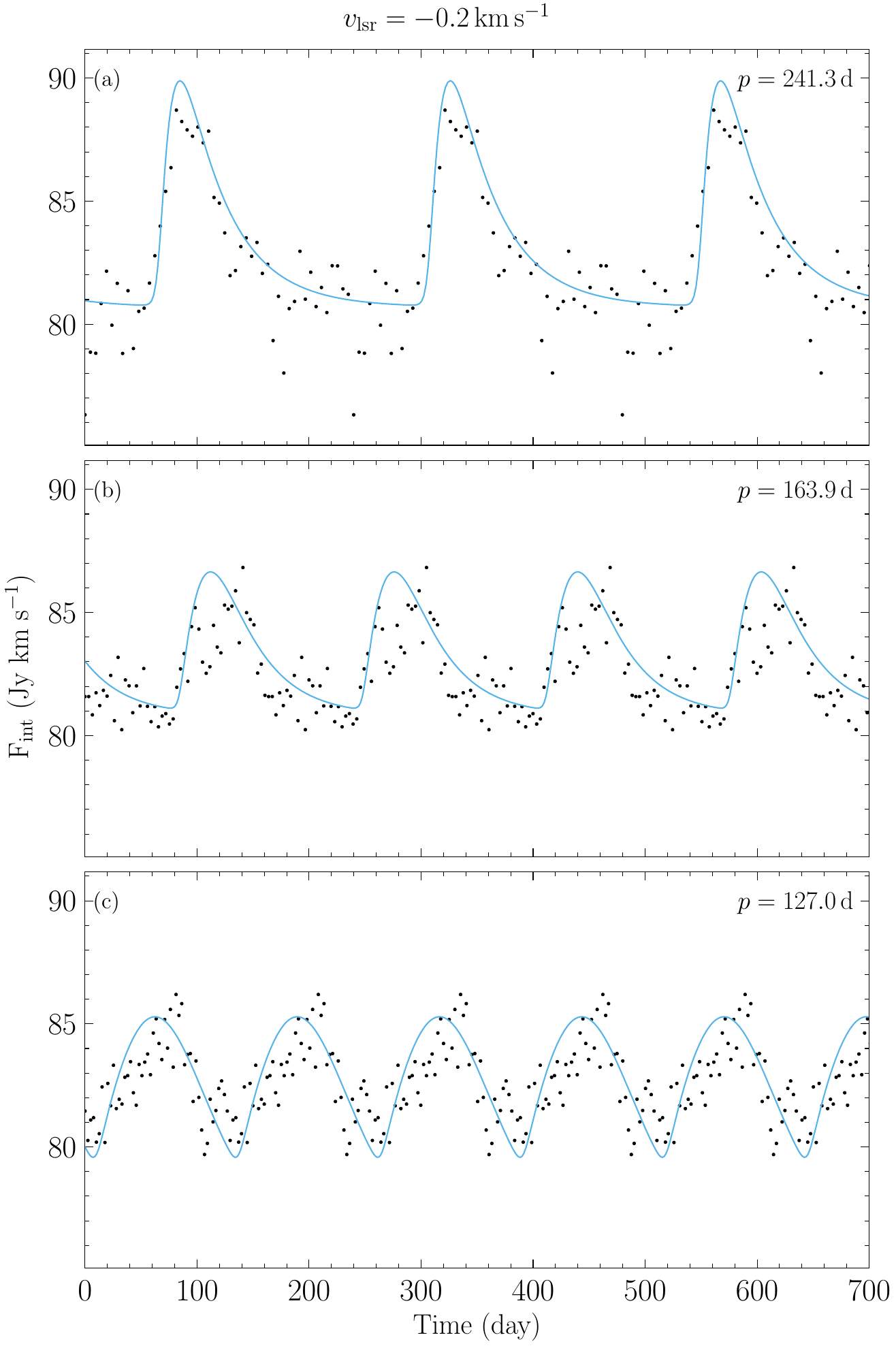}
    \caption{Phase-folded 6.7~GHz methanol flares in G9.62+0.20E at $v_{\mathrm{lsr}} = -0.2$~km~s$^{-1}$. The MBE model fits (solid blue curves) overlaid on the observational data points (black dots) are shown in each panel. The vertical axis represents the integrated flux density, while the horizontal axis shows time in days. Panels \textbf{(a)} to \textbf{(c)} correspond to the average periodicities of $p = 241.3$~d, $163.9$~d, and $127.0$~d, respectively. The MBE fits reproduce the flare profiles of all three cycles using consistent environmental parameters ($T_{1} \simeq 39.9$~d, $T_{2} \simeq 5.3$~d). The pump duration $T_{\mathrm{P}}$, pump amplitudes $\Lambda_0$ and $\Lambda_1$, and the resulting initial inverted column density $n_0L$ are as follows:  
\textbf{\textit{(a)}} $T_{\mathrm{P}} = 4.5$~d, $\Lambda_{0} = 3.8 \times 10^{-19}$~cm$^{-3}$\,s$^{-1}$, $\Lambda_{1} = 1.3 \times 10^{-19}$~cm$^{-3}$\,s$^{-1}$, yielding $n_0L = 1.0\times10^{3}$~cm$^{-2}$;  
\textbf{\textit{(b)}} $T_{\mathrm{P}} = 3.8$~d, $\Lambda_{0} = 2.2 \times 10^{-18}$~cm$^{-3}$\,s$^{-1}$, $\Lambda_{1} = 1.9 \times 10^{-19}$~cm$^{-3}$\,s$^{-1}$, yielding $n_0L = 2.6\times10^{3}$~cm$^{-2}$; \textbf{\textit{(c)}} $T_{\mathrm{P}} = 4.8$~d, $\Lambda_{0} = 3.0 \times 10^{-18}$~cm$^{-3}$\,s$^{-1}$, $\Lambda_{1} = 1.5 \times 10^{-19}$~cm$^{-3}$\,s$^{-1}$, yielding $n_0L = 8.0\times10^{3}$~cm$^{-2}$.}
    \label{fig:MBE_-0.23}
\end{figure}

\begin{figure}
    \centering
    \includegraphics[width=1.0\linewidth]{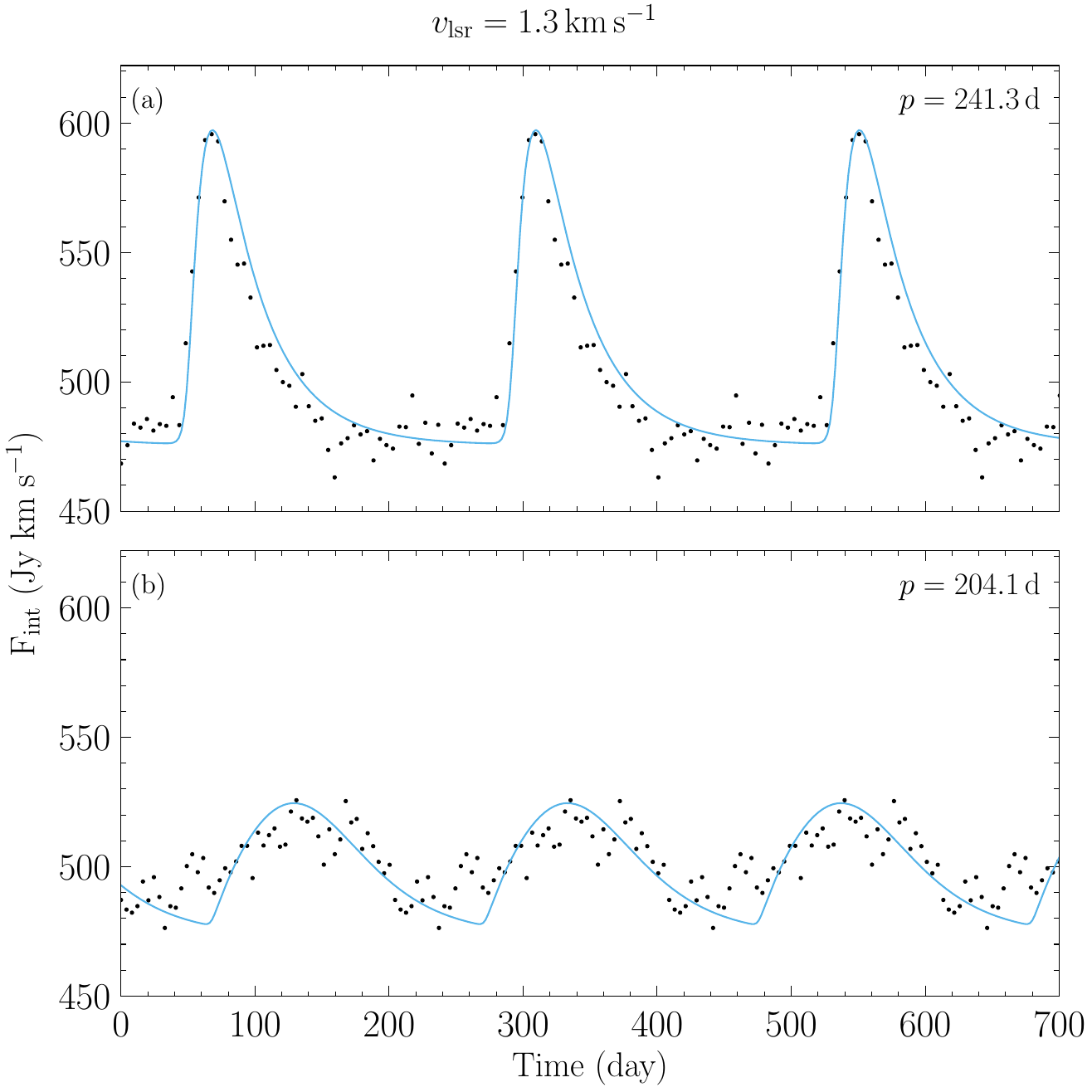}
    \caption{Same as Fig.~\ref{fig:MBE_-0.23}, but for $v_{\mathrm{lsr}} = 1.3$~km~s$^{-1}$. Panels \textbf{(a)} and \textbf{(b)} correspond to average periodicities of $p = 241.3$~d and $p = 204.1$~d, respectively. The fit parameters and the resulting initial inverted column densities are as follows: \textbf{\textit{(a)}} $T_{\mathrm{P}} = 4.5$~d, $\Lambda_0 = 4.3 \times 10^{-19}$~cm$^{-3}$\,s$^{-1}$, $\Lambda_1 = 2.9 \times 10^{-19}$~cm$^{-3}$\,s$^{-1}$, $T_1 = 37.8$~d, $T_2 = 5.1$~d, yielding $n_0L = 1.1\times10^{3}$~cm$^{-2}$; \textbf{\textit{(b)}} $T_{\mathrm{P}} = 2.8$~d, $\Lambda_0 = 3.1 \times 10^{-18}$~cm$^{-3}$\,s$^{-1}$, $\Lambda_1 = 2.5 \times 10^{-19}$~cm$^{-3}$\,s$^{-1}$, $T_1 = 41.1$~d, $T_2 = 5.2$~d, yielding $n_0L = 8.2\times10^{3}$~cm$^{-2}$.}
    \label{fig:MBE_1.32}
\end{figure}

\begin{figure}
    \centering
    \includegraphics[width=1.0\linewidth]{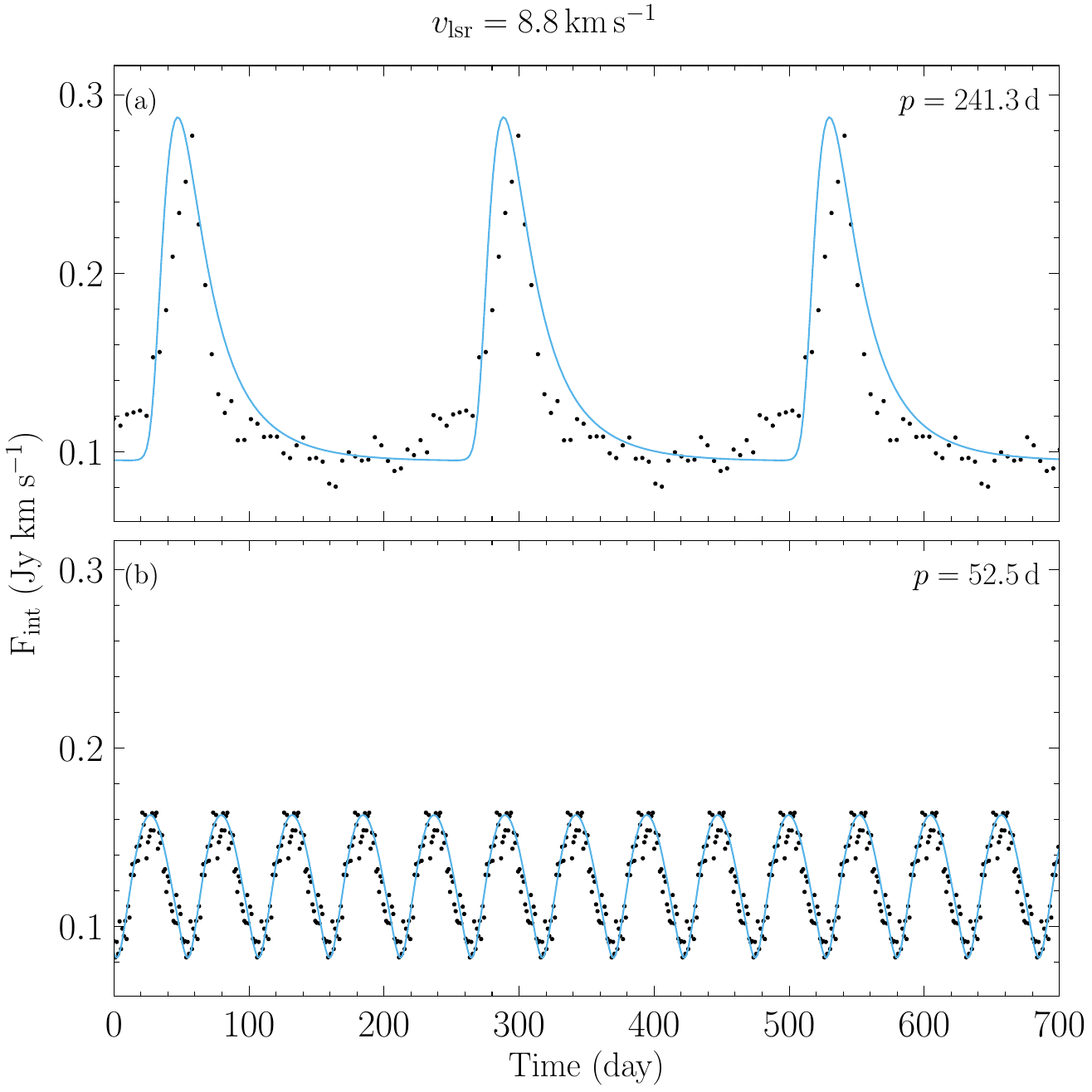}
    \caption{Same as Fig.~\ref{fig:MBE_-0.23}, but for $v_{\mathrm{lsr}} = 8.8$~km~s$^{-1}$. Panels \textbf{(a)} and \textbf{(b)} correspond to average periodicities of $p = 241.3$~d and $p = 52.5$~d, respectively. The fit parameters and the resulting initial inverted column densities are as follows: \textbf{\textit{(a)}} $T_{\mathrm{P}} = 4.5$~d, $\Lambda_0 = 4.7 \times 10^{-19}$~cm$^{-3}$\,s$^{-1}$, $\Lambda_1 = 1.5 \times 10^{-18}$~cm$^{-3}$\,s$^{-1}$, $T_1 = 35.0$~d, $T_2 = 5.0$~d, yielding $n_0L = 1.1\times10^{3}$~cm$^{-2}$; \textbf{\textit{(b)}} $T_{\mathrm{P}} = 3.4$~d, $\Lambda_0 = 1.7 \times 10^{-18}$~cm$^{-3}$\,s$^{-1}$, $\Lambda_1 = 5.3 \times 10^{-18}$~cm$^{-3}$\,s$^{-1}$, $T_1 = 40.0$~d, $T_2 = 5.4$~d, yielding $n_0L = 4.3\times10^{3}$~cm$^{-2}$.}
    \label{fig:MBE_8.76}
\end{figure}

Results for three representative velocity channels are shown in Figs.~\ref{fig:MBE_-0.23}--\ref{fig:MBE_8.76}. Each panel presents the phase-folded observational data (black points) spanning 700~d, with the corresponding MBE fits overlaid as solid blue curves. Note that the fits are produced using periodic pump pulses whose periods are matched to the corresponding average period derived from the LS periodogram, as indicated in the top-right corner of each panel.

As a representative case, Fig.~\ref{fig:MBE_-0.23} shows the $v_{\mathrm{lsr}} = -0.2$~km~s$^{-1}$ channel, which exhibits the dominant $p_{1} = 241.3$~d cycle together with the additional periodicities $p_{3} = 127.0$~d and $p_{4} = 163.9$~d. The MBE fits reproduce the flare profiles of all three cycles using consistent environmental parameters ($T_{1} \simeq 39.9$~d, $T_{2} \simeq 5.3$~d). The pump widths used in the fits lie in the range $T_{\mathrm{P}} = 3.8$--$4.5$~d, while the pump amplitudes $\Lambda_{0}$ and $\Lambda_{1}$ vary between $10^{-19}$ and $10^{-18}$~cm$^{-3}$~s$^{-1}$ depending on the period. The fits yield initial inverted column density values, $n_{0}L$, ranging from $1\times10^{3}$ to $8\times10^{3}$~cm$^{-2}$. The detailed fit parameters are listed in Table~\ref{tab:Fit_parameters} in Appendix~\ref{subsec:Parameters}.


As seen in Fig.~\ref{fig:MBE_-0.23}, panels \textbf{\textit{(a)}}--\textbf{\textit{(c)}}, the flare profiles transition from asymmetric shapes (a relatively sharp rise followed by a slow decay) to more symmetric forms as the cycle period decreases, from 241.3~d to 127.0~d. Note, the MBE fits for both asymmetric and symmetric flare profiles are produced using a symmetric pump profile (Eq.~\ref{eq:pump}).

Fig.~\ref{fig:MBE_1.32} shows the $v_{\mathrm{lsr}} = 1.3$~km~s$^{-1}$ channel, where the newly identified $p_{5} = 204.1$~d periodicity (panel~\textbf{\textit{(b)}}) appears alongside the dominant $p_{1} = 241.3$~d cycle (panel~\textbf{\textit{(a)}}). The MBE fits employ slightly different relaxation times ($T_{1} = 37.8$--41.1~d), while the dephasing timescale remains nearly constant at $T_{2} \approx 5.1$~d for both periods. The $p_{5}$ flare fit is produced using a pump duration of $T_{\mathrm{P}} = 2.8$~d, whereas that for $p_{1}$ is set to $T_{\mathrm{P}} = 4.5$~d. For consistency, these same pump durations are applied to all velocity channels exhibiting the corresponding periodicities, with $T_{\mathrm{P}} = 4.5$~d used for $p_{1} = 241.3$~d and $T_{\mathrm{P}} = 2.8$~d for $p_{5} = 204.1$~d.

The peak associated with the 204.1~d period in the LS periodogram is the weakest among the five periods discussed here (see Fig.~\ref{fig:All periodograms}), making the flare fitting for this period more uncertain, with parameters less tightly constrained and the flare shape not well defined. The pump amplitudes $\Lambda_{0}$ vary between $10^{-19}$ and $10^{-18}$~cm$^{-3}$~s$^{-1}$ across the two periods, while $\Lambda_{1} \sim 10^{-19}$~cm$^{-3}$~s$^{-1}$. The MBE models yield $n_{0}L$ values ranging from $1\times10^{3}$ to $8\times10^{3}$~cm$^{-2}$, consistent with those derived for the $v_{\mathrm{lsr}} = -0.2$~km~s$^{-1}$ channel.

For comparison, Fig.~\ref{fig:MBE_8.76} shows the $v_{\mathrm{lsr}} = 8.8$~km~s$^{-1}$ channel, which contains the two previously reported periodicities, $p_{1} = 241.3$~d and $p_{2} = 52.5$~d. The MBE fits reproduce both flare profiles using consistent environmental parameters. The models yield initial inverted column density values, $n_{0}L$, ranging from $1\times10^{3}$ to $4\times10^{3}$~cm$^{-2}$. As seen in the figure, the transition from asymmetric flares at the $\sim$241~d period to fully symmetric flares at the shorter period of $\sim$52~d is clearly evident and consistent with the behaviour observed in Fig.~\ref{fig:MBE_-0.23}, where similar changes in flare morphology occur when moving from longer to shorter periods. It should be noted that the flares at the $p_{1}$ cycle are roughly twice as long in duration as those at $p_{2} = 52.5$~d, while the pump duration in our model is varied only slightly, from $T_{\mathrm{P}} = 4.5$~d for $p_{1}$ to $T_{\mathrm{P}} = 3.4$~d for $p_{2}$, to reproduce the fits for both profiles.

Overall, these results demonstrate that the MBE framework can consistently reproduce diverse flare profiles and timescales within a single velocity component across different periods. Detailed fit parameters for all velocity components and periodicities listed in Table~\ref{tab:periods_summary} are provided in Table~\ref{tab:Fit_parameters} in Appendix~\ref{subsec:Parameters}.

\section{Discussion}\label{sec:discussion}

Our analysis of the decade-long monitoring campaign of G9.62+0.20E reveals five independent periodicities in the 6.7~GHz methanol transition: the previously established $\sim$241.3~d and 52.5~d cycles, together with three newly identified cycles at 127.0~d, 163.9~d, and 204.1~d. These periods occur across different velocity channels, with the 241.3-d and 204.1-d cycles detected in several components.

The flare profiles in G9.62+0.20E span a wide range of shapes and durations, from nearly sinusoidal (symmetric) cycles to sharp-rise/slow-decay (asymmetric) events. We show that all of these periodic flares, despite their diverse morphologies and timescales, can be consistently reproduced within the framework of the MBEs operating in the transient superradiance regime. In our modelling, each flare is driven by a narrow periodic pump whose period is closely matched to the flare period derived from its LS periodogram, while the environmental conditions (such as collisional timescales) are kept broadly uniform across all cases.

The periodic flares previously modelled using the MBE framework, either in G9.62+0.20E \citep{Rajabi2023} or in G22.356+0.066 \citep{Rashidi2025}, all exhibited asymmetric profiles similar to those shown in panel \textbf{(a)} of Figs.~\ref{fig:MBE_-0.23}--\ref{fig:MBE_8.76}. Here, we present, for the first time, MBE models for \textit{periodic} and \textit{symmetric} (sinusoidal-like) flares, such as those seen in the bottom panels of Figs.~\ref{fig:MBE_-0.23} and \ref{fig:MBE_8.76}, corresponding to the 127-d cycle at $v_{\mathrm{lsr}} = -0.2$~km~s$^{-1}$ and the 52.5-d cycle at $v_{\mathrm{lsr}} = 8.8$~km~s$^{-1}$, respectively.

Our analysis shows that flare morphology in periodic events depends sensitively on the pump-pulse period and its relation to the evolution timescales of the population inversion density and polarization within a single cycle. More specifically, when the pump period becomes comparable to, or shorter than, the time required for the system to fully relax to its pre-pulse state, the arrival of the next pump pulse interrupts this relaxation, initiating the new cycle from elevated levels of inversion and coherence. This behaviour drives $T_{\mathrm{R},0}$ to smaller values (see Table~\ref{tab:Fit_parameters}) and produces more symmetric flare profiles.

\begin{figure*}
    \centering
    \includegraphics[width=1.0\linewidth]{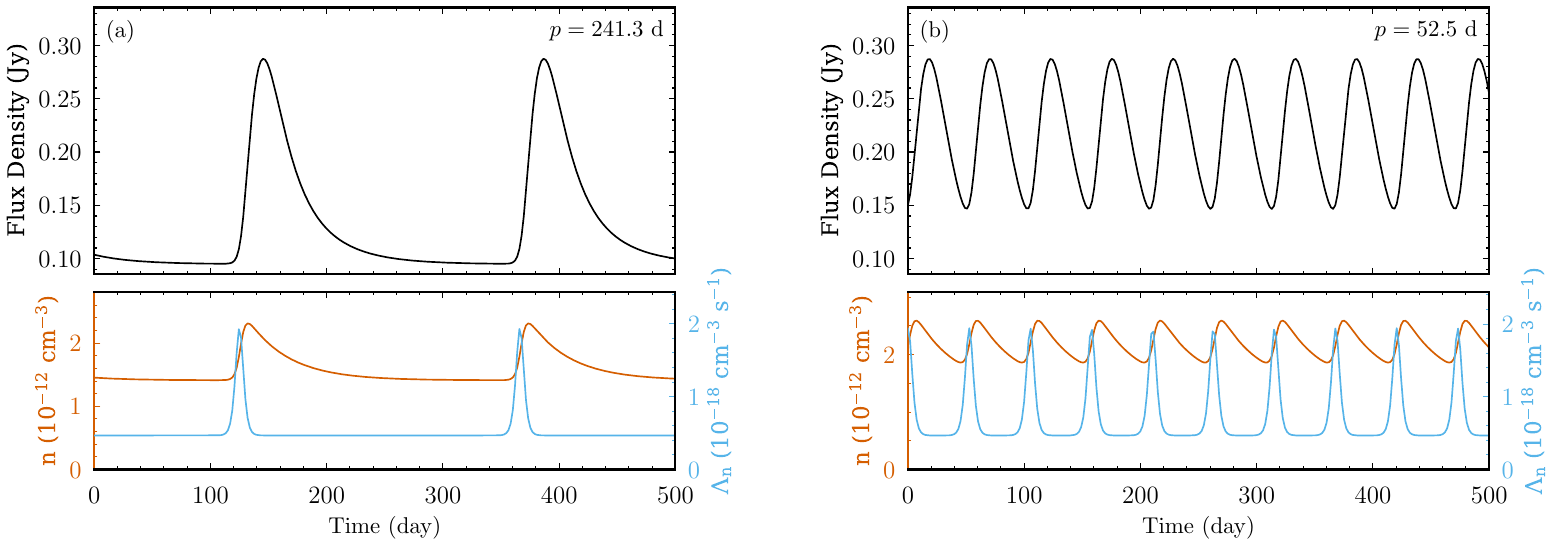}
    \caption{Effect of pump period on MBE solutions. \textbf{\textit{(a)}} \textit{Top:} Flux density from MBE solutions corresponding to the modelling of \(p_1\) for \(v_{\mathrm{lsr}}=8.8\)~km~s\(^{-1}\), using identical parameters to those in the top panel of Fig.~\ref{fig:MBE_8.76}. \textit{Bottom:} The associated population inversion density (vermilion) and periodic pump (blue). \textbf{\textit{(b)}} Similar to panel (a), but with a pump period of 52.5~d. All other parameters are identical to those of panel \textbf{\textit{(a)}} in Fig.~\ref{fig:MBE_8.76}.} 
    \label{fig:comparison}
\end{figure*}

We investigate and illustrate this behaviour in Fig.~\ref{fig:comparison} as a proof of principle. The left panels show the modelled flux density (top) and the corresponding population inversion density (bottom, left vertical axis) and pump amplitude (bottom, right vertical axis) as functions of time for a pump period of 241.3~d. For comparison, the right panels show the same quantities computed using identical fit parameters, except that the pump period is reduced to 52.5~d. Specifically, the fit parameters used in panel \textbf{(a)} of Fig.~\ref{fig:MBE_8.76} were adopted to produce the left-hand plots in Fig.~\ref{fig:comparison}. When the pump period is shortened, the population inversion does not return to its baseline level before the next pulse arrives, altering the inversion curve and yielding a smoother, more symmetric flux profile.

These results demonstrate that, in \textit{periodic flaring} events, especially when the system remains active throughout most or all of the cycle, the pump-pulse duration, period, and amplitudes jointly influence the detailed flare morphology, while the precise pump-pulse shape remains unimportant, assuming narrow excitation pulses. This behaviour differs slightly from our previous findings for asymmetric flares, where active flaring occurs only during part of the cycle and the system can return to quiescent levels between events. In Sec.~\ref{subsec:modelframework}, we stated that once the inverted column density exceeds the superradiance threshold, the cooperative response produces flares whose morphology is largely independent of the detailed pump profile, provided that the excitation pulses are narrow compared to the flare timescale. Here, we refine that statement based on our findings for periodic and symmetric flare profiles: the interplay between the pump duration, \textit{pump period}, and the system’s intrinsic evolution timescales governs the resulting flare morphology in general.

For our analysis presented here, we modelled average, phase-folded flare profiles for all identified periodicities. This approach reduced the effect of noise in the fitting procedure and was primarily motivated by the weaker cycles ($p_4$ and $p_5$), whose amplitudes are low and easily blended into the quiescent baseline of stronger events. Higher-cadence (e.g. daily) and higher-sensitivity monitoring will be essential to disentangle overlapping periodicities, confirm the stability of the weaker signals, and allow modelling of individual cycles.

In classical maser theory, the flare profile of saturated masers approximately follows the duration and shape of the pump. As a result, reproducing the observed diversity of shapes and timescales would require different pumping mechanisms, with the pump’s structure and duration playing a central role. Several scenarios within the maser picture have been proposed in this context, including colliding-wind binaries \citep{VanderWalt2009, VandenHeever2019} for sharp-rise/slow-decay flares, pulsational instabilities in young massive stars \citep{Inayoshi2013} for symmetric profiles, and orientation effects of disc–outflow systems that can alter the observed light-curve shape \citep{Morgan2021}. Our results show that, although such mechanisms may influence the astrophysical origin of the periodic driving, the flare morphologies themselves can be reproduced consistently within the unified MBE framework.

From our modelling, we obtain average values of $T_1 \sim 39$~d and $T_2 \sim 5.2$~d \footnote{Note that these average values of $T_1$ and $T_2$ correspond to all fits across 11 velocity channels and five periods, as reported in Table~\ref{tab:Fit_parameters}.}, consistent with hydrogen densities of order $10^{5}$~cm$^{-3}$ at $T = 100$~K, typical of massive star-forming regions. Both non-coherent timescales exceed the superradiance timescale $T_{R,0}$ (see Table~\ref{tab:Fit_parameters} for detailed values), ensuring that coherence can be established before collisions destroy it. In our fits, we adopted similar values of $T_1$ and $T_2$ across different velocity channels and periodicities, allowing variations of up to $\sim$9\%. Since $T_1$ and $T_2$ depend on local hydrogen densities and temperatures, this consistency indicates broadly uniform physical conditions in the regions where the flares occur. Thus, although the MBEs do not reveal the specific drivers of the multiple pump periods, they provide physically plausible constraints on the masing environment.

Several astrophysical scenarios may account for the presence of distinct periodic drivers in G9.62+0.20E. \citet{MacLeod2022} suggested either secondary pulsation modes in young massive stars \citep{Inayoshi2013} or multiple periodically varying sources in the region. Supporting the latter possibility, near-infrared observations have identified two infrared sources near G9.62+0.20E \citep{Sanna2015}.

Multi-periodic flaring is not unique to G9.62+0.20E. Two distinct periods have been reported for OH masers in the circumstellar envelopes of R~Crt and RT~Vir \citep{Etoka2001}, and velocity-dependent periodicities have been identified in G5.900$-$0.430 \citep{Tanabe2023}. Such cases indicate that multiple periodic drivers may be common in maser-hosting environments. In this context, our results show that, regardless of the specific origin of the drivers, the transient superradiance regime within the MBE framework can reproduce a broad range of flare morphologies and timescales within a single, physically consistent description. Targeted, high-resolution observations will be essential to isolate the specific drivers operating in individual sources.


\section*{Acknowledgments}
F.R.’s research is supported by the Natural Sciences and Engineering Research Council of Canada (NSERC) Discovery Grant RGPIN2024-06346. The 6.7 GHz methanol maser data obtained by the Hitachi 32-m telescope is a part of the Ibaraki 6.7 GHz Methanol Maser Monitor (iMet) program. The iMet program is partially supported by the Inter-university collaborative project 'Japanese VLBINetwork(JVN)' of NAOJ and JSPS KAKENHI grant no. JP24340034, JP21H01120, and JP21H00032 (YY).


\section*{Data Availability}
The data underlying this article will be shared on reasonable request to the corresponding author.

\bibliographystyle{mnras}
\bibliography{references}

\begin{thebibliography}{}
\makeatletter
\relax
\def\mn@urlcharsother{\let\do\@makeother \do\$\do\&\do\#\do\^\do\_\do\%\do\~}
\def\mn@doi{\begingroup\mn@urlcharsother \@ifnextchar [ {\mn@doi@}
  {\mn@doi@[]}}
\def\mn@doi@[#1]#2{\def\@tempa{#1}\ifx\@tempa\@empty \href
  {http://dx.doi.org/#2} {doi:#2}\else \href {http://dx.doi.org/#2} {#1}\fi
  \endgroup}
\def\mn@eprint#1#2{\mn@eprint@#1:#2::\@nil}
\def\mn@eprint@arXiv#1{\href {http://arxiv.org/abs/#1} {{\tt arXiv:#1}}}
\def\mn@eprint@dblp#1{\href {http://dblp.uni-trier.de/rec/bibtex/#1.xml}
  {dblp:#1}}
\def\mn@eprint@#1:#2:#3:#4\@nil{\def\@tempa {#1}\def\@tempb {#2}\def\@tempc
  {#3}\ifx \@tempc \@empty \let \@tempc \@tempb \let \@tempb \@tempa \fi \ifx
  \@tempb \@empty \def\@tempb {arXiv}\fi \@ifundefined
  {mn@eprint@\@tempb}{\@tempb:\@tempc}{\expandafter \expandafter \csname
  mn@eprint@\@tempb\endcsname \expandafter{\@tempc}}}

\bibitem[\protect\citeauthoryear{{Araya}, {Hofner}, {Goss}, {Kurtz},
  {Richards}, {Linz}, {Olmi}  \& {Sewi{\l}o}}{{Araya} et~al.}{2010}]{Araya2010}
{Araya} E.~D.,  {Hofner} P.,  {Goss} W.~M.,  {Kurtz} S.,  {Richards} A.~M.~S.,
  {Linz} H.,  {Olmi} L.,   {Sewi{\l}o} M.,  2010, \mn@doi [\apjl]
  {10.1088/2041-8205/717/2/L133}, \href
  {https://ui.adsabs.harvard.edu/abs/2010ApJ...717L.133A} {717, L133}

\bibitem[\protect\citeauthoryear{{Batrla}, {Matthews}, {Menten}  \&
  {Walmsley}}{{Batrla} et~al.}{1987}]{Batrla1987}
{Batrla} W.,  {Matthews} H.~E.,  {Menten} K.~M.,   {Walmsley} C.~M.,  1987,
  \mn@doi [\nat] {10.1038/326049a0}, \href
  {https://ui.adsabs.harvard.edu/abs/1987Natur.326...49B} {326, 49}

\bibitem[\protect\citeauthoryear{Benedict et~al.}{Benedict
  et~al.}{1996}]{Benedict1996}
Benedict M.~G.,  et~al., 1996, Super-radiance: Multiatomic Coherent Emission.
IOP Publishing Ltd

\bibitem[\protect\citeauthoryear{{Caswell}}{{Caswell}}{1998}]{Caswell1998}
{Caswell} J.~L.,  1998, \mn@doi [\mnras] {10.1046/j.1365-8711.1998.01468.x},
  \href {https://ui.adsabs.harvard.edu/abs/1998MNRAS.297..215C} {297, 215}

\bibitem[\protect\citeauthoryear{{Caswell}, {Vaile}, {Ellingsen}  \&
  {Norris}}{{Caswell} et~al.}{1995}]{Caswell1995}
{Caswell} J.~L.,  {Vaile} R.~A.,  {Ellingsen} S.~P.,   {Norris} R.~P.,  1995,
  \mn@doi [\mnras] {10.1093/mnras/274.4.1126}, \href
  {https://ui.adsabs.harvard.edu/abs/1995MNRAS.274.1126C} {274, 1126}

\bibitem[\protect\citeauthoryear{{De Buizer}, {Radomski}, {Telesco}  \&
  {Pi{\~n}a}}{{De Buizer} et~al.}{2003}]{DeBuizer2003}
{De Buizer} J.~M.,  {Radomski} J.~T.,  {Telesco} C.~M.,   {Pi{\~n}a} R.~K.,
  2003, \mn@doi [\apj] {10.1086/378949}, \href
  {https://ui.adsabs.harvard.edu/abs/2003ApJ...598.1127D} {598, 1127}

\bibitem[\protect\citeauthoryear{{Dicke}}{{Dicke}}{1954}]{Dicke1954}
{Dicke} R.~H.,  1954, \mn@doi [Physical Review] {10.1103/PhysRev.93.99}, \href
  {https://ui.adsabs.harvard.edu/abs/1954PhRv...93...99D} {93, 99}

\bibitem[\protect\citeauthoryear{{Etoka}, {B{\l}aszkiewicz}, {Szymczak}  \& {Le
  Squeren}}{{Etoka} et~al.}{2001}]{Etoka2001}
{Etoka} S.,  {B{\l}aszkiewicz} L.,  {Szymczak} M.,   {Le Squeren} A.~M.,  2001,
  \mn@doi [\aap] {10.1051/0004-6361:20011184}, \href
  {https://ui.adsabs.harvard.edu/abs/2001A&A...378..522E} {378, 522}

\bibitem[\protect\citeauthoryear{Feld \& MacGillivray}{Feld \&
  MacGillivray}{1980}]{Feld1980}
Feld M.,  MacGillivray J.,  1980, in , Coherent Nonlinear Optics.
Springer, pp 7--57

\bibitem[\protect\citeauthoryear{{Garay}, {Rodriguez}, {Moran}  \&
  {Churchwell}}{{Garay} et~al.}{1993}]{Garay1993}
{Garay} G.,  {Rodriguez} L.~F.,  {Moran} J.~M.,   {Churchwell} E.,  1993,
  \mn@doi [\apj] {10.1086/173396}, \href
  {https://ui.adsabs.harvard.edu/abs/1993ApJ...418..368G} {418, 368}

\bibitem[\protect\citeauthoryear{{Goedhart}, {Gaylard}  \& {van der
  Walt}}{{Goedhart} et~al.}{2003}]{Goedhart2003}
{Goedhart} S.,  {Gaylard} M.~J.,   {van der Walt} D.~J.,  2003, \mn@doi
  [\mnras] {10.1046/j.1365-8711.2003.06426.x}, \href
  {https://ui.adsabs.harvard.edu/abs/2003MNRAS.339L..33G} {339, L33}

\bibitem[\protect\citeauthoryear{{Goedhart}, {Gaylard}  \& {van der
  Walt}}{{Goedhart} et~al.}{2004}]{Goedhart2004}
{Goedhart} S.,  {Gaylard} M.~J.,   {van der Walt} D.~J.,  2004, \mn@doi
  [\mnras] {10.1111/j.1365-2966.2004.08340.x}, \href
  {https://ui.adsabs.harvard.edu/abs/2004MNRAS.355..553G} {355, 553}

\bibitem[\protect\citeauthoryear{{Goedhart}, {Minier}, {Gaylard}  \& {van der
  Walt}}{{Goedhart} et~al.}{2005}]{Goedhart2005}
{Goedhart} S.,  {Minier} V.,  {Gaylard} M.~J.,   {van der Walt} D.~J.,  2005,
  \mn@doi [\mnras] {10.1111/j.1365-2966.2004.08519.x}, \href
  {https://ui.adsabs.harvard.edu/abs/2005MNRAS.356..839G} {356, 839}

\bibitem[\protect\citeauthoryear{{Goedhart}, {Maswanganye}, {Gaylard}  \& {van
  der Walt}}{{Goedhart} et~al.}{2014}]{Goedhart2014}
{Goedhart} S.,  {Maswanganye} J.~P.,  {Gaylard} M.~J.,   {van der Walt} D.~J.,
  2014, \mn@doi [\mnras] {10.1093/mnras/stt2009}, \href
  {https://ui.adsabs.harvard.edu/abs/2014MNRAS.437.1808G} {437, 1808}

\bibitem[\protect\citeauthoryear{{Goedhart}, {van Rooyen}, {van der Walt},
  {Maswanganye}, {Sanna}, {MacLeod}  \& {van den Heever}}{{Goedhart}
  et~al.}{2019}]{Goedhard2019}
{Goedhart} S.,  {van Rooyen} R.,  {van der Walt} D.~J.,  {Maswanganye} J.~P.,
  {Sanna} A.,  {MacLeod} G.~C.,   {van den Heever} S.~P.,  2019, \mn@doi
  [\mnras] {10.1093/mnras/stz767}, \href
  {https://ui.adsabs.harvard.edu/abs/2019MNRAS.485.4676G} {485, 4676}

\bibitem[\protect\citeauthoryear{Gross \& Haroche}{Gross \&
  Haroche}{1982}]{Gross1982}
Gross M.,  Haroche S.,  1982, \physrep, 93, 301

\bibitem[\protect\citeauthoryear{{Hofner} \& {Churchwell}}{{Hofner} \&
  {Churchwell}}{1996}]{Hofner&Churchwell1996}
{Hofner} P.,  {Churchwell} E.,  1996, \aaps, \href
  {https://ui.adsabs.harvard.edu/abs/1996A&AS..120..283H} {120, 283}

\bibitem[\protect\citeauthoryear{{Hofner}, {Kurtz}, {Churchwell}, {Walmsley}
  \& {Cesaroni}}{{Hofner} et~al.}{1996}]{Hofner1996}
{Hofner} P.,  {Kurtz} S.,  {Churchwell} E.,  {Walmsley} C.~M.,   {Cesaroni} R.,
   1996, \mn@doi [\apj] {10.1086/176975}, \href
  {https://ui.adsabs.harvard.edu/abs/1996ApJ...460..359H} {460, 359}

\bibitem[\protect\citeauthoryear{{Houde}, {Rajabi}, {Gaensler}, {Mathews}  \&
  {Tranchant}}{{Houde} et~al.}{2019}]{Houde2019}
{Houde} M.,  {Rajabi} F.,  {Gaensler} B.~M.,  {Mathews} A.,   {Tranchant} V.,
  2019, \mn@doi [\mnras] {10.1093/mnras/sty3046}, \href
  {https://ui.adsabs.harvard.edu/abs/2019MNRAS.482.5492H} {482, 5492}

\bibitem[\protect\citeauthoryear{{Inayoshi}, {Sugiyama}, {Hosokawa}, {Motogi}
  \& {Tanaka}}{{Inayoshi} et~al.}{2013}]{Inayoshi2013}
{Inayoshi} K.,  {Sugiyama} K.,  {Hosokawa} T.,  {Motogi} K.,   {Tanaka} K.
  E.~I.,  2013, \mn@doi [\apjl] {10.1088/2041-8205/769/2/L20}, \href
  {https://ui.adsabs.harvard.edu/abs/2013ApJ...769L..20I} {769, L20}

\bibitem[\protect\citeauthoryear{{Jurkevich}}{{Jurkevich}}{1971}]{Jurkevich1971}
{Jurkevich} I.,  1971, \mn@doi [\apss] {10.1007/BF00656321}, \href
  {https://ui.adsabs.harvard.edu/abs/1971Ap&SS..13..154J} {13, 154}

\bibitem[\protect\citeauthoryear{{Kidger}, {Takalo}  \& {Sillanpaa}}{{Kidger}
  et~al.}{1992}]{kidger1992}
{Kidger} M.,  {Takalo} L.,   {Sillanpaa} A.,  1992, \aap, \href
  {https://ui.adsabs.harvard.edu/abs/1992A&A...264...32K} {264, 32}

\bibitem[\protect\citeauthoryear{Lomb}{Lomb}{1976}]{Lomb1976}
Lomb N.~R.,  1976, \mn@doi [Astrophysics and Space Science]
  {10.1007/BF00648343}, 39, 447–462

\bibitem[\protect\citeauthoryear{{MacLeod} et~al.,}{{MacLeod}
  et~al.}{2022}]{MacLeod2022}
{MacLeod} G.~C.,  et~al., 2022, \mn@doi [\mnras] {10.1093/mnrasl/slac083},
  \href {https://ui.adsabs.harvard.edu/abs/2022MNRAS.516L..96M} {516, L96}

\bibitem[\protect\citeauthoryear{{Menten}}{{Menten}}{1991}]{Menten1991}
{Menten} K.~M.,  1991, \mn@doi [\apjl] {10.1086/186177}, \href
  {https://ui.adsabs.harvard.edu/abs/1991ApJ...380L..75M} {380, L75}

\bibitem[\protect\citeauthoryear{{Morgan}, {van der Walt}, {Chibueze}  \&
  {Zhang}}{{Morgan} et~al.}{2021}]{Morgan2021}
{Morgan} J.,  {van der Walt} D.~J.,  {Chibueze} J.~O.,   {Zhang} Q.,  2021,
  \mn@doi [\mnras] {10.1093/mnras/stab2185}, \href
  {https://ui.adsabs.harvard.edu/abs/2021MNRAS.507.1138M} {507, 1138}

\bibitem[\protect\citeauthoryear{{Parfenov} \& {Sobolev}}{{Parfenov} \&
  {Sobolev}}{2014}]{Parfenov2014}
{Parfenov} S.~Y.,  {Sobolev} A.~M.,  2014, \mn@doi [\mnras]
  {10.1093/mnras/stu1481}, \href
  {https://ui.adsabs.harvard.edu/abs/2014MNRAS.444..620P} {444, 620}

\bibitem[\protect\citeauthoryear{{Rajabi} \& {Houde}}{{Rajabi} \&
  {Houde}}{2016a}]{Rajabi2016a}
{Rajabi} F.,  {Houde} M.,  2016a, \mn@doi [\apj] {10.3847/0004-637X/826/2/216},
  \href {https://ui.adsabs.harvard.edu/abs/2016ApJ...826..216R} {826, 216}

\bibitem[\protect\citeauthoryear{{Rajabi} \& {Houde}}{{Rajabi} \&
  {Houde}}{2016b}]{Rajabi2016b}
{Rajabi} F.,  {Houde} M.,  2016b, \mn@doi [\apj] {10.3847/0004-637X/828/1/57},
  \href {https://ui.adsabs.harvard.edu/abs/2016ApJ...828...57R} {828, 57}

\bibitem[\protect\citeauthoryear{{Rajabi} \& {Houde}}{{Rajabi} \&
  {Houde}}{2017}]{Rajabi2017}
{Rajabi} F.,  {Houde} M.,  2017, \mn@doi [Science Advances]
  {10.1126/sciadv.1601858}, \href
  {https://ui.adsabs.harvard.edu/abs/2017SciA....3E1858R} {3, e1601858}

\bibitem[\protect\citeauthoryear{{Rajabi} \& {Houde}}{{Rajabi} \&
  {Houde}}{2020}]{Rajabi2020}
{Rajabi} F.,  {Houde} M.,  2020, \mn@doi [\mnras] {10.1093/mnras/staa1067},
  \href {https://ui.adsabs.harvard.edu/abs/2020MNRAS.494.5194R} {494, 5194}

\bibitem[\protect\citeauthoryear{{Rajabi}, {Houde}, {Bartkiewicz}, {Olech},
  {Szymczak}  \& {Wolak}}{{Rajabi} et~al.}{2019}]{Rajabi2019}
{Rajabi} F.,  {Houde} M.,  {Bartkiewicz} A.,  {Olech} M.,  {Szymczak} M.,
  {Wolak} P.,  2019, \mn@doi [\mnras] {10.1093/mnras/stz074}, \href
  {https://ui.adsabs.harvard.edu/abs/2019MNRAS.484.1590R} {484, 1590}

\bibitem[\protect\citeauthoryear{{Rajabi}, {Houde}, {MacLeod}, {Goedhart},
  {Tanabe}, {van den Heever}, {Wyenberg}  \& {Yonekura}}{{Rajabi}
  et~al.}{2023}]{Rajabi2023}
{Rajabi} F.,  {Houde} M.,  {MacLeod} G.~C.,  {Goedhart} S.,  {Tanabe} Y.,  {van
  den Heever} S.~P.,  {Wyenberg} C.~M.,   {Yonekura} Y.,  2023, \mn@doi
  [\mnras] {10.1093/mnras/stad2671}, \href
  {https://ui.adsabs.harvard.edu/abs/2023MNRAS.526..443R} {526, 443}

\bibitem[\protect\citeauthoryear{{Rashidi}, {Anari}, {Bartkiewicz}, {Wolak},
  {Szymczak}  \& {Rajabi}}{{Rashidi} et~al.}{2025}]{Rashidi2025}
{Rashidi} T.,  {Anari} V.,  {Bartkiewicz} A.,  {Wolak} P.,  {Szymczak} M.,
  {Rajabi} F.,  2025, \mn@doi [\mnras] {10.1093/mnrasl/slaf061}, \href
  {https://ui.adsabs.harvard.edu/abs/2025MNRAS.542L..12R} {542, L12}

\bibitem[\protect\citeauthoryear{{Roberts}, {Lehar}  \& {Dreher}}{{Roberts}
  et~al.}{1987}]{Roberts1987}
{Roberts} D.~H.,  {Lehar} J.,   {Dreher} J.~W.,  1987, \mn@doi [\aj]
  {10.1086/114383}, \href
  {https://ui.adsabs.harvard.edu/abs/1987AJ.....93..968R} {93, 968}

\bibitem[\protect\citeauthoryear{{Sanna}, {Reid}, {Moscadelli}, {Dame},
  {Menten}, {Brunthaler}, {Zheng}  \& {Xu}}{{Sanna} et~al.}{2009}]{Sanna2009}
{Sanna} A.,  {Reid} M.~J.,  {Moscadelli} L.,  {Dame} T.~M.,  {Menten} K.~M.,
  {Brunthaler} A.,  {Zheng} X.~W.,   {Xu} Y.,  2009, \mn@doi [\apj]
  {10.1088/0004-637X/706/1/464}, \href
  {https://ui.adsabs.harvard.edu/abs/2009ApJ...706..464S} {706, 464}

\bibitem[\protect\citeauthoryear{{Sanna} et~al.,}{{Sanna}
  et~al.}{2015}]{Sanna2015}
{Sanna} A.,  et~al., 2015, \mn@doi [\apjl] {10.1088/2041-8205/804/1/L2}, \href
  {https://ui.adsabs.harvard.edu/abs/2015ApJ...804L...2S} {804, L2}

\bibitem[\protect\citeauthoryear{{Scargle}}{{Scargle}}{1982}]{Scargle1982}
{Scargle} J.~D.,  1982, \mn@doi [\apj] {10.1086/160554}, \href
  {https://ui.adsabs.harvard.edu/abs/1982ApJ...263..835S} {263, 835}

\bibitem[\protect\citeauthoryear{Stellingwerf}{Stellingwerf}{1978}]{Stellingwerf1978}
Stellingwerf R.~F.,  1978, Astrophysical Journal, Part 1, vol. 224, Sept. 15,
  1978, p. 953-960., 224, 953

\bibitem[\protect\citeauthoryear{{Tanabe}, {Yonekura}  \& {MacLeod}}{{Tanabe}
  et~al.}{2023}]{Tanabe2023}
{Tanabe} Y.,  {Yonekura} Y.,   {MacLeod} G.~C.,  2023, \mn@doi [\pasj]
  {10.1093/pasj/psad002}, \href
  {https://ui.adsabs.harvard.edu/abs/2023PASJ...75..351T} {75, 351}

\bibitem[\protect\citeauthoryear{{Testi}, {Felli}, {Persi}  \& {Roth}}{{Testi}
  et~al.}{1998}]{Testi1998a}
{Testi} L.,  {Felli} M.,  {Persi} P.,   {Roth} M.,  1998, \mn@doi [\aaps]
  {10.1051/aas:1998403}, \href
  {https://ui.adsabs.harvard.edu/abs/1998A&AS..129..495T} {129, 495}

\bibitem[\protect\citeauthoryear{{Testi}, {Hofner}, {Kurtz}  \&
  {Rupen}}{{Testi} et~al.}{2000}]{Testi2000}
{Testi} L.,  {Hofner} P.,  {Kurtz} S.,   {Rupen} M.,  2000, \mn@doi [\aap]
  {10.48550/arXiv.astro-ph/0006211}, \href
  {https://ui.adsabs.harvard.edu/abs/2000A&A...359L...5T} {359, L5}

\bibitem[\protect\citeauthoryear{{Yonekura} et~al.,}{{Yonekura}
  et~al.}{2016}]{Yonekura2016}
{Yonekura} Y.,  et~al., 2016, \mn@doi [\pasj] {10.1093/pasj/psw045}, \href
  {https://ui.adsabs.harvard.edu/abs/2016PASJ...68...74Y} {68, 74}

\bibitem[\protect\citeauthoryear{{van den Heever}, {van der Walt}, {Pittard}
  \& {Hoare}}{{van den Heever} et~al.}{2019}]{VandenHeever2019}
{van den Heever} S.~P.,  {van der Walt} D.~J.,  {Pittard} J.~M.,   {Hoare}
  M.~G.,  2019, \mn@doi [\mnras] {10.1093/mnras/stz576}, \href
  {https://ui.adsabs.harvard.edu/abs/2019MNRAS.485.2759V} {485, 2759}

\bibitem[\protect\citeauthoryear{{van der Walt}, {Goedhart}  \& {Gaylard}}{{van
  der Walt} et~al.}{2009}]{VanderWalt2009}
{van der Walt} D.~J.,  {Goedhart} S.,   {Gaylard} M.~J.,  2009, \mn@doi
  [\mnras] {10.1111/j.1365-2966.2009.15147.x}, \href
  {https://ui.adsabs.harvard.edu/abs/2009MNRAS.398..961V} {398, 961}

\makeatother
\end{thebibliography}

\appendix

\section{Maxwell-Bloch Equations}\label{subsec:MBEs}

The dynamics of a sample of two-level atoms or molecules interacting with a radiation field are described by the Maxwell--Bloch equations (MBEs). Within the Rotating Wave Approximation (RWA) and the Slowly Varying Envelope Approximation (SVEA) \citep{Gross1982,Benedict1996}, the MBEs can be written as  

\begin{equation} \label{eq:MBE-n}
    \frac{\partial n'}{\partial \tau} = \frac{i}{\hbar} \left( P^+ E^+ - P^- E^- \right) - \frac{n'}{T_1} + \Lambda_n ,
\end{equation}

\begin{equation} \label{eq:MBE-P}
    \frac{\partial P^+}{\partial \tau} = \frac{2i d^2}{\hbar} E^- n' - \frac{P^+}{T_2},
\end{equation}

\begin{equation} \label{eq:MBE-E}
    \frac{\partial E^+}{\partial z} = \frac{i \omega_0}{2 \epsilon_0 c} P^-,
\end{equation}
where $\tau = t - z/c$ is the retarded time, $d = |\mathbf{d}|$ is the transition dipole moment, $\omega_0 = ck$ is the transition frequency, and $\epsilon_0$ is the vacuum permittivity. The population inversion density is $n = 2n'$, $P$ is the amplitude of the induced atomic or molecular polarization, and $E$ is the electric-field amplitude. The corresponding polarization and field vectors are expressed as  

\begin{equation}
\begin{split}
    \mathbf{P}^{\pm}(z, \tau) &= P^{\pm}(z, \tau)\, e^{\pm i \omega_0 \tau} \,\boldsymbol{\epsilon}_d, \\
    \mathbf{E}^{\pm}(z, \tau) &= E^{\pm}(z, \tau)\, e^{\mp i \omega_0 \tau} \,\boldsymbol{\epsilon}_d,
\end{split}
\end{equation}
where $\boldsymbol{\epsilon}_d = \mathbf{d}/d$ is the unit polarization vector. The superscripts $``+"$ and $``-"$ denote the positive- and negative-frequency components of the field, with $E^- = (E^+)^*$. For the polarization, $P^+$ corresponds to the transition from the lower to upper level, while $P^-$ corresponds to the transition from the upper to lower level, with $P^- = (P^+)^*$.  

The phenomenological terms $-n'/T_1$ and $-P^+/T_2$ represent, respectively, non-coherent population relaxation and polarization dephasing. The effective pump $\Lambda_n$ in Eq.~\ref{eq:MBE-n} maintains periodic inversion in the sample. The temporal evolution of the system is initiated by internal fluctuations in the polarization and population inversion.

To ensure that the one-dimensional MBEs adequately model the system, we impose a Fresnel number of unity on the samples, which sets the cross-sectional area to $A = \lambda L$, where $\lambda$ is the radiation wavelength and $L$ is the sample length. The equations are solved numerically using a fourth-order Runge–Kutta method. The radiation field amplitude at the end-fire of the sample ($z = L$) is then used to compute the output intensity,  
\begin{equation}
    I = \tfrac{1}{2} c \epsilon_0 |E|^2 ,
\end{equation}
which is fitted to the observational data.

\section{Period analysis}\label{subsec:periodogram-analysis}

We applied three independent methods of period detection to our datasets: the Lomb–Scargle (LS) periodogram \citep{Lomb1976, Scargle1982}, the epoch-folding method described by \citet{Stellingwerf1978}, and the Jurkevich method \citep{Jurkevich1971}.

We computed LS power spectra for trial periods of 10--300~d and found a dominant signal at $p_{1} = 241.3 \pm 2.3$~d, consistent with \citet{Goedhart2003}. We also identified a secondary period at $p_{2} = 52.5 \pm 0.3$~d, consistent with \citet{MacLeod2022}, along with additional periods  at 
$p_{3} = 127.0 \pm 1.6$~d, $p_{4} = 163.9 \pm 2.9$~d, and $p_{5} = 204.1 \pm 1.5$~d.

To assess peak significance, we adopted an empirical detection threshold derived from noise simulations. Specifically, we averaged the maximum power from 1000 LS periodograms generated from synthetic light curves with flux values drawn from the observational noise range (\(\pm 0.3\)~Jy). This procedure yielded an LS power threshold of 0.007; peaks below this level are consistent with pure noise. For peaks above this level, we evaluated whether they represent genuine signals, harmonics of real signals, or artifacts introduced by the finite time window of the LS analysis.

A finite time window introduces artifacts, most notably sidelobes around strong peaks and their harmonics. To test whether such features correspond to genuine periodicities, we applied the CLEAN algorithm to the time-series data \citep{Roberts1987}. CLEAN iteratively subtracts the dominant peak and its aliases from the periodogram while suppressing noise, yielding spectra in which statistically significant peaks can be identified more robustly. The 204.1~d period could, in principle, be misidentified as a harmonic of the 241.3~d signal; however, its persistence in the CLEANed periodograms supports its reality. For clarity, the CLEANed periodograms are not shown, but were used to confirm the dominant periods reported.

Having established the LS periodogram results, we applied epoch folding to further test the periodicities. This method computes the phase of each observation time $t_i$ by folding over a trial period $P$:
\begin{equation}
    \phi_i = \frac{t_i}{P} - \left[\frac{t_i}{P} \right],
\end{equation}
where the square bracket denotes the integer part of the enclosed value. Trial periods spanning 10--300~d were tested, matching the LS search range. The resulting phases and flux densities were grouped into 70 phase bins. We then applied the Phase Dispersion Minimization (PDM) statistic \citep{Stellingwerf1978} to identify the period that minimizes phase scatter.

As a third test, we applied the Jurkevich method. As with epoch folding, we folded and binned the data; however, instead of comparing bin means with the overall mean (as in PDM), this method evaluates the variance within each bin. The true period corresponds to the minimum variance. The significance of the result is quantified as
\begin{equation}
    f = \frac{1 - V_m^2}{V_m^2},
\end{equation}
where $V_m^2$ is the variance within bin $m$ \citep{kidger1992}. Values of $f > 0.5$ indicate strong periodicity; for $P_1$ to $P_5$ we obtained $f > 0.9$ in all cases.

\section{Table of Parameters}\label{subsec:Parameters}

\begin{table*}
\centering
\makebox[\textwidth]{\parbox{0.9\textwidth}{
\caption{For each velocity channel, this table lists the measured period(s) together with the parameters that yield the best MBE fits to the average pulses. These parameters include the pump amplitudes \(\Lambda_0\) and \(\Lambda_1\), the pump duration \(T_{\}mathrm{P}}\), and the non-coherent timescales \(T_1\) and \(T_2\). From these values, the initial superradiance timescale \(T_{R,0}\) and their corresponding inverted column density \(n_0 L\) are calculated and also shown.}
\label{tab:Fit_parameters}
}}
\begin{tabular}{ |p{1.5cm}|p{1.5cm}|p{1cm}|p{1cm}|p{1cm}|p{2cm}|p{2cm}|p{1cm}|p{1.5cm}|}

\hline
\( v_{\mathrm{lsr}}~(\mathrm{km~s}^{-1}) \) & \( p \)~(d) & \( T_1 \)~(d) & \( T_2 \)~(d) & \( T_{\mathrm{P}} \)~(d) & \(\Lambda_0 \)~(cm\(^{-3}\)~s\(^{-1}\))  &\(\Lambda_1\)~(cm\(^{-3}\)~s\(^{-1}\)) & \( T_{R,0} \)~(d) & \(n_0L~(\mathrm{cm^{-2}})\) \\
\hline
\hline
-0.8 & 204.1 & 42.0 & 5.6 & 2.8 & $2.4\times 10^{-18}$ & $1.9\times 10^{-19}$ & 0.5& $6.5\times 10^{3}$ \\
\hline
-0.2 & 241.3 & 41.7 & 5.2 & 4.5 & $3.8\times 10^{-19}$ & $1.3\times 10^{-19}$ & 3.0& $1.0\times 10^{3}$ \\
& 163.9 & 36.7 & 5.5 & 3.8 & $2.2\times 10^{-18}$ & $1.9\times 10^{-19}$ & 1.2& $2.6\times 10^{3}$ \\
& 127.0 & 41.2 & 5.1 & 4.8 & $3.0\times 10^{-18}$ & $1.5\times 10^{-19}$ & 0.4& $8.0\times 10^{3}$ \\
\hline
0.4 & 241.3 & 40.0 & 5.4 & 4.5 & $5.9\times 10^{-19}$ & $3.3\times 10^{-19}$ & 2.8& $1.1\times 10^{3}$ \\
& 204.1 & 36.5 & 5.6 & 2.8 & $3.2\times 10^{-18}$ & $2.4\times 10^{-19}$ & 0.4& $7.6\times 10^{3}$ \\
\hline
1.3 & 241.3 & 37.8 & 5.1 & 4.5 & $4.3\times 10^{-19}$ & $2.9\times 10^{-19}$ & 2.9& $1.1\times 10^{3}$ \\
& 204.1 & 41.1 & 5.2 & 2.8 & $3.1\times 10^{-18}$ & $2.5\times 10^{-19}$ & 0.4& $8.2\times 10^{3}$ \\
\hline
3.3 & 241.3 & 38.3 & 5.0 & 4.5 & $4.4\times 10^{-19}$ & $2.8\times 10^{-19}$ & 3.0& $1.0\times 10^{3}$ \\
& 204.1 & 38.4 & 5.4 & 2.8 & $4.2\times 10^{-18}$ & $3.2\times 10^{-19}$ & 0.3& $1.0\times 10^{4}$ \\
\hline
4.1 & 241.3 & 41.3 & 5.2 & 4.5 & $1.0\times 10^{-18}$ & $3.3\times 10^{-19}$ & 2.8& $1.1\times 10^{3}$ \\
& 204.1 & 41.9 & 5.7 & 2.8 & $2.2\times 10^{-18}$ & $2.1\times 10^{-19}$ & 0.5& $6.0\times 10^{3}$ \\
\hline
5.0 & 241.3 & 35.0 & 4.9 & 4.5 & $4.3\times 10^{-19}$ & $8.5\times 10^{-19}$ & 3.2& $9.7\times 10^{2}$ \\
& 204.1 & 40.6 & 5.1 & 2.8 & $3.9\times 10^{-18}$ & $1.2\times 10^{-18}$ & 0.3& $1.0\times 10^{4}$ \\
\hline
5.4 & 204.1 & 39.2 & 5.0 & 2.8 & $3.6\times 10^{-18}$ & $1.0\times 10^{-19}$ & 0.3& $9.1\times 10^{3}$ \\
\hline
6.5 & 204.1 & 40.0 & 4.9 & 2.8 & $3.4\times 10^{-18}$ & $1.4\times 10^{-19}$ & 0.3& $8.8\times 10^{3}$ \\
\hline
8.1 & 241.3 & 35.0 & 5.0 & 4.5 & $4.1\times 10^{-19}$ & $3.7\times 10^{-18}$ & 3.3& $9.3\times 10^{2}$ \\
\hline
8.8 & 241.3 & 35.0 & 5.0 & 4.5 & $4.7\times 10^{-19}$ & $1.5\times 10^{-18}$ & 2.9& $1.1\times 10^{3}$ \\
& 52.5 & 40.0 & 5.4 & 3.4 & $1.7\times 10^{-18}$ & $5.3\times 10^{-18}$ & 0.7& $4.3\times 10^{3}$ \\

\hline
    
\end{tabular}
\end{table*}

\label{lastpage}
\end{document}